\documentclass{article} 
\usepackage{iclr2021_conference,times}

\usepackage{amsmath,amsfonts,bm}

\def\eqref#1{equation~\ref{#1}}

\def\1{\bm{1}}

\DeclareMathAlphabet{\mathsfit}{\encodingdefault}{\sfdefault}{m}{sl}
\SetMathAlphabet{\mathsfit}{bold}{\encodingdefault}{\sfdefault}{bx}{n}

\usepackage{hyperref}
\usepackage{url}

\usepackage{booktabs}       
\usepackage{amsfonts}       
\usepackage{nicefrac}       
\usepackage{microtype}      
\usepackage{graphicx} 
\usepackage{float} 
\usepackage{subfigure} 
\usepackage{amsmath}
\usepackage{wrapfig,lipsum,booktabs}

\usepackage{enumitem}

\title{HiFiSinger: Towards High-Fidelity Neural Singing Voice Synthesis}

\author{Jiawei Chen, ~Xu Tan\thanks{Corresponding author}, ~Jian Luan, ~Tao Qin, ~Tie-Yan Liu   \\
Microsoft STC Asia \& Microsoft Research Asia  \\
\texttt{\{t-jiawch,xuta,jianluan,taoqin,tyliu\}@microsoft.com} \\
}

\iclrfinalcopy 
\begin{document}

\maketitle

\begin{abstract}
High-fidelity singing voices usually require higher sampling rate (e.g., 48kHz, compared with 16kHz or 24kHz in speaking voices) with large range of frequency to convey expression and emotion. However, higher sampling rate causes the wider frequency band and longer waveform sequences and throws challenges for singing modeling in both frequency and time domains in singing voice synthesis (SVS). Conventional SVS systems that adopt moderate sampling rate (e.g., 16kHz or 24kHz) cannot well address the above challenges. In this paper, we develop HiFiSinger, an SVS system towards high-fidelity singing voice using 48kHz sampling rate. HiFiSinger consists of a FastSpeech based neural acoustic model and a Parallel WaveGAN based neural vocoder to ensure fast training and inference and also high voice quality. To tackle the difficulty of singing modeling caused by high sampling rate (wider frequency band and longer waveform), we introduce multi-scale adversarial training in both the acoustic model and vocoder to improve singing modeling. Specifically, 1) To handle the larger range of frequencies caused by higher sampling rate (e.g., 48kHz vs. 24kHz), we propose a novel sub-frequency GAN (SF-GAN) on mel-spectrogram generation, which splits the full 80-dimensional mel-frequency into multiple sub-bands (e.g. low, middle and high frequency bands) and models each sub-band with a separate discriminator. 2) To model longer waveform sequences caused by higher sampling rate, we propose a multi-length GAN (ML-GAN) for waveform generation to model different lengths of waveform sequences with separate discriminators. 3) We also introduce several additional designs and findings in HiFiSinger that are crucial for high-fidelity voices, such as adding F0 (pitch) and V/UV (voiced/unvoiced flag) as acoustic features, choosing an appropriate window/hop size for mel-spectrogram, and increasing the receptive field in vocoder for long vowel modeling in singing voices. Experiment results show that HiFiSinger synthesizes high-fidelity singing voices with much higher quality: 0.32/0.44 MOS gain over 48kHz/24kHz baseline and 0.83 MOS gain over previous SVS systems. Audio samples are available at \url{https://speechresearch.github.io/hifisinger/}.
\end{abstract}

\section{Introduction}
Singing voice synthesis (SVS) aims to synthesize high-quality and expressive singing voices based on musical score information, and attracts a lot of attention in both industry and academia (especially in the machine learning and speech signal processing community)~\citep{umbert2015expression,nishimura2016singing,blaauw2017neural,nakamura2019singing,hono2019singing,chandna2019wgansing,lee2019adversarially,lu2020xiaoicesing,blaauw2020sequence,gu2020bytesing,ren2020deepsinger}. Singing voice synthesis shares similar pipeline with text to speech synthesis, and has achieved rapid progress~\citep{blaauw2017neural,nakamura2019singing,lee2019adversarially,blaauw2020sequence,gu2020bytesing} with the techniques developed in text to speech synthesis~\citep{shen2018natural,ren2019fastspeech,ren2020fastspeech,yamamoto2020parallel}. 

Most previous works on SVS~\citep{lee2019adversarially,gu2020bytesing} adopt the same sampling rate (e.g., 16kHz or 24kHz) as used in text to speech, where the frequency bands or sampling data points are not enough to convey expression and emotion as in high-fidelity singing voices. However, simply increasing the sampling rate will cause several challenges in singing modeling.
First, the audio with higher sampling rate contains wider and higher frequency bands\footnote{According to Nyquist-Shannon sampling theorem~\citep{millette2013heisenberg}, a sampling rate $f_s$ can cover the frequency band up to $f_s / 2$. Therefore, the frequency band for the audio with 48kHz sampling rate spans from 0$\sim$24kHz while 0$\sim$12kHz for 24kHz sampling rate. The additional high frequency band 12$\sim$24kHz increases the difficulty of modeling since high-frequency signals are more complicated and less predictive.}, which throws challenges when predicting these frequency spectrums in acoustic model.
Second, the audio with higher sampling rate contains longer waveform points and much fine-grained fluctuations in a fixed period of time\footnote{For example, a 1 second audio waveform contains 48,000 sampling points when sampling rate is 48kHz.}, which also increases the difficulty of vocoder modeling in time domain. As a consequence, even if some previous works~\citep{hono2019singing,chandna2019wgansing,wu2019synthesising,nakamura2020fast,lu2020xiaoicesing} adopt higher sampling rate (e.g. 44.1kHz or 48kHz), they either leverage coarse-grained MFCC~\citep{zheng2001comparison} as acoustic features in slow autoregressive neural vocoder~\citep{oord2016wavenet}, or use non-neural vocoder such as Griffin-Lim~\citep{griffin1984signal} and WORLD~\citep{morise2016world} to generate waveform, which do not fully exploit the potential of high sampling rate and thus cannot yield good voice quality.

In this paper, we develop HiFiSinger, an SVS system towards high-fidelity singing voices. HiFiSinger adopts FastSpeech~\citep{ren2019fastspeech} as the acoustic model and Parallel WaveGAN~\citep{yamamoto2020parallel} as the vocoder since they are popular in speech synthesis~\citep{hayashi2020espnet,ren2020fastspeech,blaauw2020sequence,lu2020xiaoicesing} to ensure fast training and inference speed and also high quality.
To address the challenges of high sampling rate in singing modeling (wider frequency band and longer waveform), we design multi-scale adversarial training on both acoustic model and vocoder, and introduce several additional systematic designs and findings that are crucial to improve singing modeling:
\begin{itemize}[leftmargin=*]
\item To handle larger range of frequencies caused by higher sampling rate (e.g., 0$\sim$24kHz in 48kHz vs. 0$\sim$12kHz in 24kHz) and model high-frequency details for high-fidelity singing voices, we propose a novel sub-frequency GAN (SF-GAN) on mel-spectrogram generation, which splits the full 80-dimensional mel-frequency into multiple sub-bands (e.g., low, middle and high frequency bands) and models each sub-band with a separate discriminator.
\item To model longer waveform caused by higher sampling rate, we propose a multi-length GAN (ML-GAN) on waveform generation, which randomly crops different lengths of waveform sequence and model them with separate discriminators. As a result, singing voices can be modeled in different length granularities to avoid the issues (e.g., glitches and vibrations) occurred in a single discriminator with a fixed length of waveform sequence.
\item We further introduce several designs and findings in HiFiSinger that are important to achieve high-fidelity synthesis: 1) Besides mel-spectrogram, we add pitch (fundamental frequency, F0) and V/UV (voiced/unvoiced flag) as acoustic features to better model singing voices; 2) We carefully study the window and hop size in acoustic features and choose an appropriate value to better align with the range of pitches in singing voices and also trade off the modeling difficulty between acoustic model and vocoder; 3) We increase the receptive field in vocoder to cover long vowel in singing voices.
\end{itemize}

We conduct experiments on our internal singing voice synthesis datasets that contain 11 hours high-fidelity singing recordings with 48kHz sampling rate. Experiment results demonstrate the advantages of our developed HiFiSinger over previous singing voice synthesis system. Further ablation studies verify the effectiveness of each design in HiFiSinger to generate high-fidelity voices.

\section{Background}
In this section, we brieﬂy introduce the background of this work, including the comparison between singing voice synthesis (SVS) and text to speech (TTS), the challenges of high fidelity singing voice synthesis.

\paragraph{SVS vs. TTS}
Text to speech (TTS) aims to synthesize speech voice from a given text, which has evolved quickly from early concatenative synthesis \citep{hunt1996unit}, statistical parametric synthesis~\citep{wu2016merlin,li2018emphasis}, to neural network based parametric synthesis~\citep{arik2017deep}, and to currently end-to-end neural models. The end-to-end models directly map input text or phonetic characters to output speech, which greatly simplifies the training pipeline and reduces the requirements for linguistic and acoustic knowledge. Popular end-to-end TTS systems include FastSpeech~\citep{ren2019fastspeech,ren2020fastspeech}, Tacotron 2~\citep{shen2018natural}, etc. With the rapid development, TTS has been applied to various scenarios and has been the basic technology in singing voice synthesis (SVS)~\citep{chandna2019wgansing,lu2020xiaoicesing,ren2020deepsinger,gu2020bytesing}. However, SVS has distinct features compared with TTS, since SVS needs more information (note pitch and note duration) in addition to the given lyric (text) to synthesize singing voices with wide range of pitches, long vowel durations. Furthermore, singing voices focus more on expression and emotion rather than content as in speaking voices, which requires higher sampling rate than speaking voices to ensure high-fidelity voices and thus throws great challenges for singing modeling. 

\paragraph{High-Fidelity SVS}
Singing voices usually leverage high sampling rate to convey high-fidelity expression. For example, popular music websites such as Spotify, Apple Music, SoundCloud, QQ Music and NetEase Music all use high sampling rate (44.1kHz or higher). However, high sampling rate increases the difficulty of singing modeling: 1) high sampling rate causes wider spectrum band in frequency domain, where different frequency bands with distinctive characteristics make it hard for acoustic model; 2) high sampling rate causes longer waveform in a fixed period of time, where more sampling points and finer-grained fluctuations make it difficult for vocoder. Most previous works on SVS usually adopt 16kHz or 24kHz sampling rate as used in TTS. There indeed exist some works using 44.1kHz or 48kHz sampling rate~\citep{hono2019singing,chandna2019wgansing,wu2019synthesising,nakamura2020fast,lu2020xiaoicesing}. However, they either leverage coarse-grained MFCC~\citep{zheng2001comparison} as acoustic features in slow autoregressive neural vocoder~\citep{oord2016wavenet}, or use non-neural vocoder such as Griffin-Lim~\citep{griffin1984signal} and WORLD~\citep{morise2016world} to generate waveform, which cannot fully exploit the potential of high sampling rate and thus cannot yield good voice quality.


\section{Method}
\label{sec_method}
In this section, we first introduce the overall architecture of HiFiSinger, and then describe the specific designs to address the distinctive challenges caused by high sampling rate in singing modeling, including sub-frequency GAN (SF-GAN) for wider frequency band, multi-length GAN (ML-GAN) for longer waveform, and several systematic designs and findings that are important for high quality singing voices.

\subsection{System Overview}

\begin{figure}[t]
  \centering
  \includegraphics[scale=0.73]{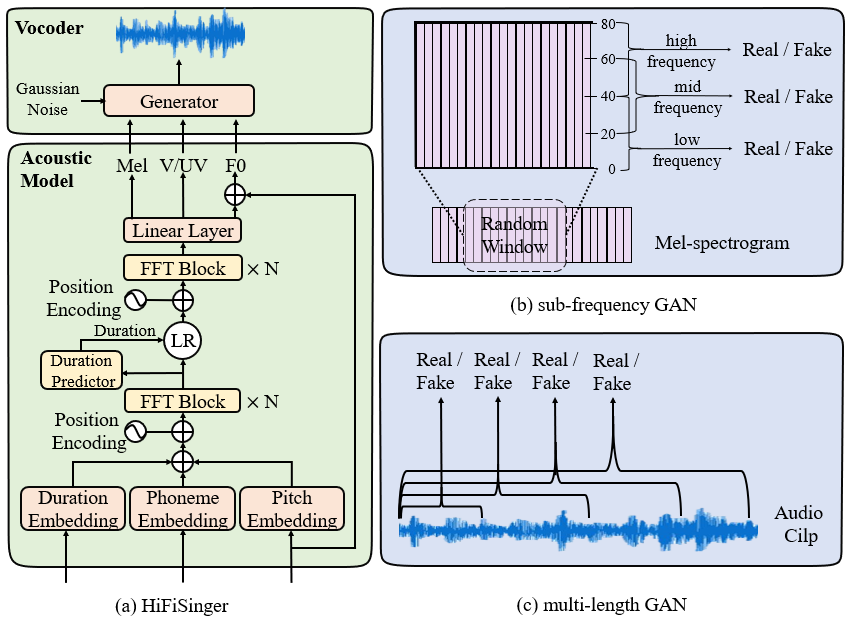}\\
  \caption{(a)The overall architecture of HiFiSinger, which consist of a parallel acoustic model and a parallel vocoder. (b) The sub-frequency GAN on mel-spectrogram. (c) The multi-length GAN on waveform.}
  \label{fig:overview}
\end{figure}

A typical SVS system consists of an acoustic model to convert the music score into acoustic features, and a vocoder to generate audio waveform from acoustic features. As illustrated in Figure \ref{fig:overview}(a), to ensure high-quality synthesized voice and fast training and inference speed, HiFiSinger consists of an acoustic model based on FastSpeech~\citep{ren2019fastspeech, ren2020fastspeech} and a vocoder based on Parallel WaveGAN~\citep{yamamoto2020parallel}, both of which are non-autoregressive generation models. We introduce the details of the data input and model structure as follows.

\paragraph{Music Score Input} In order to generate high quality singing voice with good pronunciation, tone, rhythm and timbre, we use music score that contains lyrics, note pitch and note duration as the input of acoustic model. Specifically, we process the music score as follows: 
1) We convert the character (e.g., Chinese) or syllable (e.g., English) in lyrics into phoneme using grapheme-to-phoneme conversion~\citep{taylor2005hidden, sun2019token}. 
2) We convert each note into pitch ID according to the MIDI standard\footnote{\url{https://www.midi.org/}. For example, the pitch ID corresponds to note C4 is 60, about 262Hz.}. 
3) We quantize the note duration according to the music tempo and then convert it to represent the number of frames of mel-spectrograms\footnote{For example, given a tempo 120, there are 120 beats in one minute and one beat in 0.5 second. For a time signature 4/4, a quarter note has a duration of 0.5 second. If the hop size of mel-spectrogram is 5ms, then a quarter note corresponds to 100 frames.}. We repeat the pitch and duration ID to match the length of phoneme in each character or syllable. Thus, the musical score input can be represented a sequence $x \in \mathbb{R} ^{N \times 3}$, where $N$ is the number of phonemes, and 3 represents the three IDs for phoneme, pitch and duration, which are embedded in dense vectors, added together as the input of acoustic model.

\paragraph{Acoustic Model and Vocoder} The acoustic model is built on FastSpeech, which uses a feed-forward Transformer (FFT) block~\citep{vaswani2017attention} as the basic structure of the encoder and decoder. Since the singing voices do not exactly follow the duration in the music score, we need to explicitly predict the duration for natural and expressive singing voice. We use a duration predictor to predict how many frames of mel-spectrograms that each phoneme corresponds to in the singing voice, and expand the phoneme hidden sequence to match the length of mel-spectrograms. The vocoder consists of a parallel generator as used in Parallel WaveGAN.

\subsection{Modeling Wide Frequency with SF-GAN}
\label{sec_method_sfgan}
In order to generate highly expressive and high-fidelity singing voice, larger sampling rate is needed to cover more high-frequency details, which has wider frequency bands in mel-spectrograms. As a consequence, it increases the difficulty of mel-spectrogram modeling since wide frequency bands are full of diverse and complicated patterns, especially in the additional high frequency band 12$\sim$24kHz. A natural idea is to use larger mel bins (e.g., 120 vs. 80) for mel-spectrogram representation, where mel bin 80$\sim$120 is used to cover additional high-frequency information. However, we have tried in experiments and found no obvious improvements in voice quality. Actually, the key is not to increase the bins of mel-spectrogram\footnote{Simply increasing mel bins will not bring much information unless increasing the STFT (short-time Fourier transformation) filter size at the same time. However, since there is a trade-off between the resolutions of frequency and time~\citep{landau1961prolate}, increasing the frequency bins equals to increase the frequency resolution, which requires the sacrifice of time resolution (related to window size). But according to our careful experiment studies, the optimal window size is 20ms. Using other window sizes will cause worse voice quality.}, but how to better model the diverse frequency details in a wide range of frequency band. 

A common practice is to leverage generative adversarial network (GAN) to improve the mel-spectrogram predictions and avoid over-smoothing problem caused by mean square error (L2) loss or mean absolute error (L1) loss. However, a single discriminator is difficult to cover the diverse patterns across different frequency bands. Therefore, we propose a sub-frequency GAN (SF-GAN) to model the singing audio with high sampling rate, which leverages multiple discriminators on top of the acoustic model for adversarial training of mel-spectrograms, as shown in Figure \ref{fig:overview}(b). We split the mel-spectrograms into multiple frequency bands and handle each frequency band with a separate discriminator. 
The discriminator of each frequency band focuses on guiding the sub-spectrogram of the corresponding frequency band to be less over-smoothing and closer to ground truth. The formulation of SF-GAN is shown in Equation~\ref{eq_mf_gan_g} and~\ref{eq_mf_gan_d}:
\begin{flalign}
  \begin{split}
     \min_{G_{\text{am}}} \mathbb{E}_{x}[\sum_{f \in \{\text{low}, \text{mid}, \text{high}\}}(1-D_f(G_{\text{am}}(x))^2)],
\label{eq_mf_gan_g}
  \end{split}\\  \begin{split}
     \min_{D_f} {\mathbb{E}}_{y} [(1-D_f(y))^2] + \mathbb{E}_{x} [D_f(G_{\text{am}}(x)], \forall f \in \{\text{low}, \text{mid}, \text{high}\},
\label{eq_mf_gan_d}
  \end{split}
\end{flalign}
where the GAN loss follows LS-GAN~\citep{mao2017least} considering it is popular in speech, $x$ and $y$ represent music score input and mel-spectrogram output respectively, $G_{\text{am}}$ represents the acoustic model and $D_f$ represents the discriminator for frequency band $f$. For example, for an 80-dimensional mel-spectrogram, we split it into low, medium and high frequency band, where the lowest 40-dimension (0 to 40) as low-frequency, the middle 40-dimension (20 to 60) as mid-frequency, and the highest 40-dimension (40 to 80) as high-frequency, and each frequency band has overlap with adjacent bands.

\subsection{Modeling Long Waveform with ML-GAN}
\label{sec_method_mlgan}
For a high sampling rate audio, it not only means that a wider frequency band in frequency domain, but also a longer waveform in time domain, which means more fine-grained and complicated fluctuations in fixed range of time. Previous vocoders~\citep{yamamoto2020parallel} usually adopt a single discriminator to distinguish the entire audio clip, which cannot well handle the fluctuation patterns in different time ranges in the long waveform sequence. Therefore, we design a multi-length GAN (ML-GAN) in HiFiSinger, as shown in Figure~\ref{fig:overview}(c), which uses multiple discriminators to distinguish the sampling points in different lengths. The formulation of ML-GAN is shown in Equation~\ref{eq_ml_gan_g} and ~\ref{eq_ml_gan_d}:
 \begin{flalign}
  \begin{split}
     \min_{G_{\text{voc}}} \mathbb{E}_{y}[\sum_{t \in (0, len(w))}(1-D_t(G_{\text{voc}}(y))^2)],
\label{eq_ml_gan_g}
  \end{split}\\
  \begin{split}
     \min_{D_t} {\mathbb{E}}_{w} [(1-D_t(w))^2] + \mathbb{E}_{y} [D_t(G_{\text{voc}}(y)], \forall t \in (0, len(w)),
\label{eq_ml_gan_d}
  \end{split}
\end{flalign}
where the GAN loss follows LS-GAN~\citep{mao2017least}, $y$ and $w$ represent acoustic feature input (including mel-spectrogram, F0 and V/UV) and waveform output respectively, $G_{\text{voc}}$ represents the vocoder and $D_t$ represents the discriminator for different time length $t$. The benefits of ML-GAN are twofold: 1) it reduces the difficulty of longer waveform modeling (caused by high sampling rate) by modeling shorter waveform sequence; 2) it can better capture the dynamic phoneme duration (too long or too short) in singing voices via modeling different lengths of waveform sequences.

\subsection{Other Systematic Designs}
\label{sec_method_designs}

Compared with speaking voices, singing voices have a larger range of pitches and phoneme durations, which also throws challenges in singing modeling. Therefore, we further introduce some systematic designs and findings in HiFiSinger that are crucial to improve the voice quality, including using pitch and U/UV as additional acoustic features, carefully studying window size and hop size to trade off between acoustic model and vocoder, and increasing the receptive field in vocoder to better model long vowels in singing voices. We describe them as follows:
\begin{itemize}[leftmargin=*]
    \item Pitch and V/UV. Singing voices heavily rely on pitch for voice quality. Therefore, besides mel-spectrograms, our acoustic model also predicts pitch where we use the original note pitch in music score as shortcut input to let the model focus on learning the residual pitch value, as shwon in Figure~\ref{fig:overview}(a). Besides, we also make a voiced/unvoiced (V/UV) flag to help correct the pitch values and avoid electronic noise as shown in the experiment section. The vocoder takes the mel-spectrogram, pitch and V/UV as input to generate waveform with better quality. 
    \item Window/Hop size. There are two considerations in the choices of window and hop size: 1) The window size of mel-spectrogram during short-time Fourier transformation needs careful study since larger pitch prefers smaller window size while smaller pitch prefers larger window size\footnote{Usually, the window size should cover 2$\sim$8 times of the period of the fundamental frequency~\citep{juvela2016high, kawahara2006straight}. For example, for a pitch of 100Hz (which is common in speaking voices), the period is 10ms and the window size should fall between 20$\sim$80 ms. As we can see, the window size in speaking voices is usually set to 50ms~\citep{shen2018natural,ren2019fastspeech}, which falls into this range.}. The pitch in singing voices is usually higher (sometimes maybe lower) than speaking voices, and thus the window size needs to be smaller than that in speaking voices. 2) A smaller hop size will cause the acoustic features more fine-grained and longer in sequence length, which is more difficult for acoustic model to predict but beneficial to vocoder due to more fine-grained input. On the other hand, a larger hop size will ease the acoustic model training but will increase the difficulty of vocoder training. After careful study, we set window size as 20ms and hop size as 5ms (under a relationship of 4:1 following the common practice~\citep{shen2018natural,ren2019fastspeech}).
    \item Large receptive field. Furthermore, unlike speaking voices, the duration in the music note and corresponding lyric may vary a lot, causing a larger range of phoneme duration (usually longer on vowels). To better model the large range of duration, we use a larger kernel size in the vocoder to enlarge the receptive field to cover such long vowels.
\end{itemize}

\section{Experiments and Results}

In this section, we first describe the experimental setup, and then report the experiment results, including audio quality, ablation study and analysis of our proposed system.

\subsection{Experimental Setup}

\paragraph{Datasets}
Our singing datasets contains Chinese Mandarin pop songs collected from a female singer, who sings with the accompaniment in a professional recording studio. All the singing recordings are sampled at 48kHz, quantized with 16 bits and split into pieces between 3 and 10 seconds. The final datasets contain 6817 pieces, about 11 hours of data. We randomly choose 340 pieces for validation and 340 for test. When extracting mel-spectrogram features, the window size and hop size are set to 20ms and 5ms and the number of mel bins are set to 80. We extract the F0 and V/UV label from the singing audio\footnote{We extract F0 using Parselmouth from~\url{https://github.com/YannickJadoul/Parselmouth}, and set a voiced label if F0 \textless 3, otherwise unvoiced.} and get the phoneme duration label (used in the duration predictor) with HMM-based forced alignment~\citep{sjolander2003hmm}. Both the mel-spectrogram and F0 features are normalized to have zero mean and unit variance before training, respectively.

\paragraph{Model Conﬁguration}
The backbone of the acoustic model is based on FastSpeech, where both the encoder and decoder consist of 6 FFT blocks. In each block, the hidden size of self-attention is set to 384 and the kernel width/input size/output size in the two-layer 1D-convolution are set to 3/384/1536 and 1/1536/384 respectively. On top of the last FFT block, a linear layer is used to generate the 80-dimensional mel-spectrogram, a one-dimensional F0 (float value)  and a one-dimensional V/UV (0-1 value) as the acoustic features. The basic structure of the vocoder is based on WaveNet, where 10 non-causal dilated 1D-convolution layers with dilations of 1, 2, 4, ..., 512 are stacked 3 times. The channel size for dilations, residual blocks, and skip-connections are 64, 128, and 64, respectively. Specially, the kernel size of each 1D-convolution layer is set to 13 to model high sampling rate audio, as described in Section~\ref{sec_method_designs}. 

We then describe the discriminator in SF-GAN (acoustic model) and ML-GAN (vocoder) respectively. SF-GAN consists of three discriminators for low (0$\sim$40), middle (20$\sim$60) and high (40$\sim$80) frequency mel bins respectively. At the same time, in each frequency band, the corresponding discriminator does not judge the whole generated or real mel-spectrogram sequence, but just a random sub-sampling fragments with different length of random windows, similar to~\cite{binkowski2019high}, which has been demonstrated to have a data augmentation effect and also reduces the computational complexity. All discriminators share the same model structure but different model parameters, each with three 2D-convolution layers followed by a Leaky ReLU activation function and a linear projection for final output. ML-GAN consists of four discriminators for 0.25s, 0.5s, 0.75s and 1.0s of randomly sampled waveform sequence. Each discriminator consist of 10 non-causal dilated 1D-convolutions layers with the Leaky ReLU activation function whose dilations increase linearly starting from the first layer. The number of channels are the same as the generator of vocoder and the kernel size is set to 9. 

\paragraph{Training and Inference}

We train the acoustic model and vocoder separately. The acoustic model is trained for 60k steps with minibatch size of 32 using Adam optimizer ($\beta_1=0.9, \beta_2=0.98, \epsilon=10^{-9}$) and the same learning rate schedule in \citet{ren2019fastspeech}. The vocoder is trained for 400k steps with minibatch size of 4 using RAdam \citep{liu2019variance} optimizer. The initial learning rate is set to 0.0001, and was reduced by half for every 200k steps. Note that the discriminators are turned on starting from 10k steps in SF-GAN and 100k steps in ML-GAN to warm up the generators in acoustic model and vocoder. During training, we use the ground-truth label of the phoneme duration in acoustic model and the ground-truth mel-spectrogram, F0 and V/UV as in the input of vocoder, while during inference we use the corresponding predicted values.

\subsection{Audio Quality}

\begin{wraptable}{r}{0.50\textwidth}
\vspace{-1.0cm}
\centering
\small
\caption{The MOS with 95\% conﬁdence intervals. 48kHz sampling rate is used unless otherwise stated.}
\label{mos}
\begin{tabular}{lc}
\\ \toprule[1.5pt]
\textbf{Method}            & \textbf{MOS}        \\ 
\midrule[1pt]
Recording         & $4.03 \pm  0.06$ \\
Recording (24kHz)  & $3.70 \pm  0.08$  \\
\midrule
XiaoiceSing~\citep{lu2020xiaoicesing}       & $2.93 \pm 0.06$  \\
\midrule
Baseline (24kHz)          & $3.32 \pm 0.09$  \\
Baseline (24kHz upsample)       & $3.38 \pm 0.08$   \\
Baseline                  & $3.44 \pm 0.08$  \\
\midrule
HiFiSinger (24kHz) & $3.47 \pm  0.06$ \\
HiFiSinger         & $3.76 \pm  0.06$ \\
\bottomrule[1.5pt]
\end{tabular}
 \vspace{-0.2cm}
\end{wraptable}

To verify the effectiveness of the proposed HiFisinger system, we conduct the MOS (mean opinion score) evaluation on the test set (we randomly choose 100 pieces from the test set for evaluation) to measure the quality of the synthesized singing voices. Each audio is listened by at least 20 judgers. We mainly compare HiFiSinger with the following settings and systems: 1) Recording, the original singing recordings; 2) Recording (24kHz), the original singing recordings downsampled to 24kHz; 3) XiaoiceSing~\citep{lu2020xiaoicesing}, a previous SVS system that also adopts 48kHz sampling rate but leverages WORLD vocoder; 4) Baseline (24kHz), a baseline SVS system that uses the basic model backbone of HiFiSinger (FastSpeech based acoustic model and Parallel WaveGAN based vocoder) but without any of our improvements in HiFiSinger (SF-GAN, ML-GAN and other systematic improvements as described in Section~\ref{sec_method}), and only uses 24kHz sampling rate; 5) Baseline (24kHz upsample), waveform generated by Baseline (24kHz) is upsampled to 48kHz; 6) Baseline (48kHz), the same baseline system as in 4) but uses 48kHz sampling rate; 7) HiFiSinger (24kHz), our proposed HiFiSinger system but uses 24kHz sampling rate; 8) HiFiSinger, our final HiFiSinger system with 48kHz sampling rate\footnote{The audio samples are available at ~\url{https://speechresearch.github.io/hifisinger/}}.


Experiments results are shown in Table~\ref{mos}. We have several observations: 1)  HiFiSinger outperforms XiaoiceSing and Baseline by 0.83 MOS and 0.32 MOS respectively at the sampling rate of 48kHz, which demonstrates the effectiveness of HiFiSinger for singing voices with high sampling rate. 2) When increasing the audio sampling rate from 24kHz to 48kHz, Baseline has only 0.12 MOS gain (3.44 vs. 3.32) while HiFiSinger has 0.29 MOS gain (3.76 vs. 3.47), which also demonstrates the potential of HiFiSinger for high sampling rate. 3) HiFiSinger with 48kHz sampling rate even achieves higher MOS score than the 24kHz recordings, and only has 0.27 MOS gap to the 48kHz recordings, which verifies the high-fidelity voices synthesized by HiFiSinger.

\subsection{Ablation Studies}

We conduct ablation studies to verify the effectiveness of several components in HiFiSinger, including 1) sub-frequency GAN (SF-GAN), 2) multi-length GAN (ML-GAN), 3) pitch and V/UV, 4) window and hop size, 5) large receptive field. We mainly conduct CMOS evaluation to compare two different settings, where each of the randomly chosen 100 evaluation pieces in the test set are listened by 20 judgers.

\paragraph{SF-GAN}

\begin{wraptable}{r}{0.50\textwidth}
\vspace{-0.6cm}
\centering
\small
\caption{The CMOS results for SF-GAN, where $n$ SF-GAN represents there are $n$ disciminators handling different frequency bands in SF-GAN.}
\label{cmos1}
\begin{tabular}{lc}
\\ \toprule[1.5pt]
\textbf{System}                   & \textbf{CMOS}   \\ 
\midrule[1pt]
HiFiSinger (default 3 SF-GAN)     & 0      \\
\midrule
HiFiSinger with 0 SF-GAN          & -0.22 \\
HiFiSinger with 1 SF-GAN          & -0.28 \\
HiFiSinger with 5 SF-GAN          & -0.06 \\
\bottomrule[1.5pt]
\end{tabular}
\vspace{-0.2cm}
\end{wraptable}

We explore the performance when varying the number of discriminators in SF-GAN (described in Section~\ref{sec_method_sfgan}). We make the total number of parameters of the discriminators in different settings comparable (e.g., the total parameters of 3 discriminator is same as that of 1 discriminator)\footnote{We conduct experiments to make the parameters in 1 SF-GAN to be 1/3 times of that in 3 SF-GAN, which causes even worse voice quality. Therefore, to be fair, we keep the total parameters of each setting the same.}. From Table \ref{cmos1}, it can be seen that HiFiSinger with 3 SF-GAN (default) outperforms other settings with 1) 0 SF-GAN (without any discriminator), which shows the advantages of adversarial training; 2) 1 SF-GAN, which shows that a single discriminator cannot handle the complicated and diverse patterns in low, middle and high frequency band; 3) 5 SF-GAN (the mel bins of the 5 sub-frequency band is $0\sim26, 13\sim39, 26\sim52, 39\sim65, 52\sim80$), which shows that using more discriminators slightly hurt the quality. Therefore, we choose 3 discriminators as the default setting. As discussed in Section~\ref{sec_method_sfgan}, another possible idea is to increase the number of mel bins to cover more high-frequency bands. Therefore, we conduct experiments to evaluate the voice quality when increasing the number of mel bins from 80 to 120 (both using a single discriminator), and find there is only 0.02 CMOS gain, which demonstrates that simply increasing the number of mel bins cannot well model the diverse frequency details over a wider band.

\begin{figure}[h]
  \centering
  \includegraphics[scale=0.65]{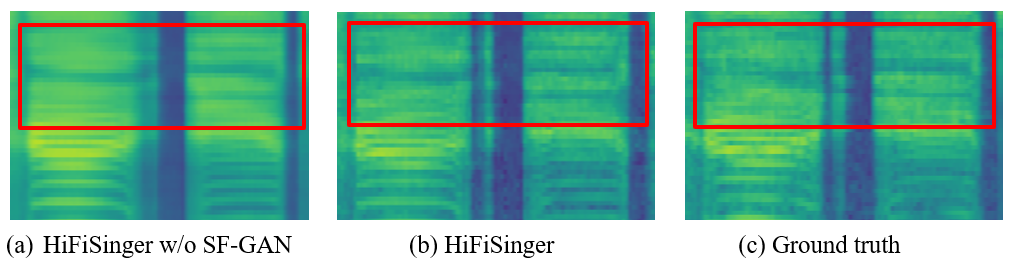}\\
  \caption{The mel-spectrogram comparisons for SF-GAN.}
  \label{fig:mel_comparison}
\end{figure}

Figure \ref{fig:mel_comparison} shows the generated mel-spectrograms of HiFiSinger, HiFiSinger without discriminator, and the ground truth. It can be seen that HiFiSinger without discriminator generates over-smoothing mel-spectrogram, and after adding SF-GAN, the me-spectrogram have more high frequency details and are closer to the ground truth.

\begin{wraptable}{r}{0.50\textwidth}
\vspace{-1.2cm}
\centering
\small
\caption{The CMOS results for ML-GAN and single length GAN.}
\label{cmos2b}
\begin{tabular}{lc}
\\ \toprule[1.5pt]
\textbf{System}                   & \textbf{CMOS}   \\ 
\midrule[1pt]
HiFiSinger with ML-GAN  & 0  \\
\midrule
HiFiSinger with 0.25s length & -0.21 \\
HiFiSinger with 0.50s length & -0.38 \\
HiFiSinger with 0.75s length & -0.15 \\
HiFiSinger with 1.00s length & -0.12 \\
\bottomrule[1.5pt]
\end{tabular}
\vspace{-0.2cm}
\end{wraptable}

\begin{figure}[htbp]
  \centering
  \includegraphics[scale=0.65]{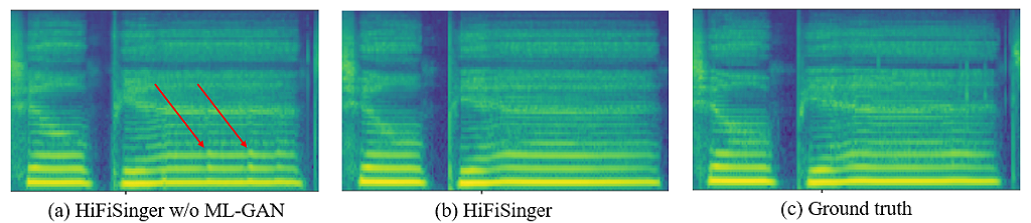}\\
  \caption{The mel-spectrogram comparisons for ML-GAN.}
  \label{fig:mel_comparison_b}
\end{figure}

\paragraph{ML-GAN}
We further study the effectiveness of ML-GAN in modeling long waveform caused by high sampling rate. As shown in Table~\ref{cmos2b}, it can be seen that only using a single discriminator on a certain length of waveform sequence (0.25s, 0.50s, 0.75s or 1.00s length, i.e., w/o ML-GAN) performs worse than HiFiSinger with ML-GAN (multiple discriminators on 0.25/0.5/0.75/1s)\footnote{According to our case analyses, a single discriminator with a small length results in expressive voice but electronic noise, while a single discriminator with a big length results in less expressive voice but few electronic noise, and all single length usually has glitches and vibrations in long vowel. However, ML-GAN can combine the advantages of discriminators with different lengths of waveform and avoid these issues.}. Figure~\ref{fig:mel_comparison_b} shows a sample case. It can be seen that there is a glitch in the long vowel generated by HiFiSinger w/o ML-GAN (use a single length of 1s discriminator), while HiFiSinger can generate stable long vowel similar to the ground truth, thanks to the finer granularity modeling of long vowel by multi-length discriminators.

\paragraph{Other System Designs}
Next, we study the effectiveness of other system designs to improve the high fidelity singing quality, including adding pitch and V/UV as the vocoder input, window and hop size choice and larger receptive field.


\begin{figure}[htbp]
  \centering
  \includegraphics[scale=0.64]{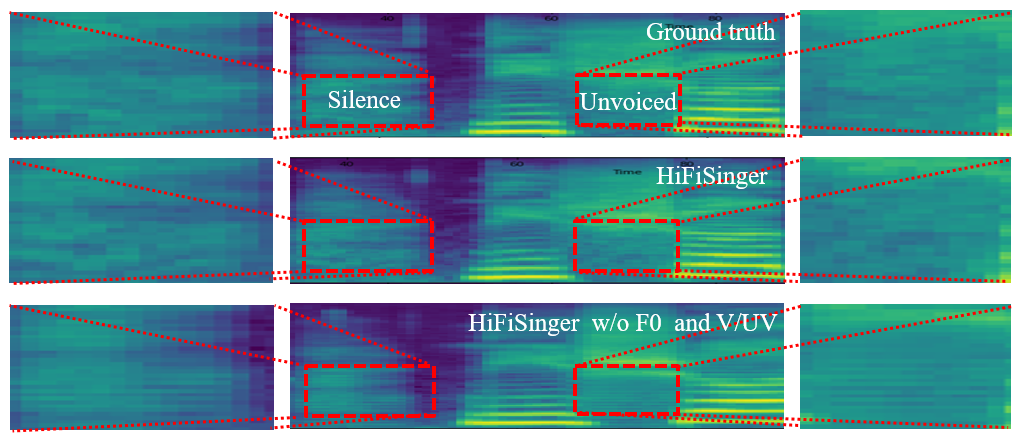}\\
  \caption{The mel-spectrograms comparisons of HiFiSinger w/ and w/o pitch and V/UV, where ``silence'' and ``unvoiced'' represent the silence frames and unvoiced frames.}
  \label{uv_input}
\end{figure}


\emph{Pitch and V/UV} F0 and V/UV can help the vocoder model the pitch and differentiate the speech with voiced and unvoiced frames. We conduct the CMOS evaluation on pitch and V/UV, as shown in Table \ref{cmos3}. Removing pitch and V/UV from the vocoder input results in a 0.34 CMOS drop and only removing V/UV causes a 0.28 CMOS drop, which demonstrates the effectiveness of pitch and V/UV. 
As shown in Figure~\ref{uv_input}, removing F0 and V/UV make the unvoiced part (including silence and unvoiced frames) less informative and over-smoothing, which causes electronic noise according to our experimental observations. Besides, pitch can make the vocoder more controllable and more robust to larger pitch range. We show in the demo page that we can change the pitch (increase or decrease several semitones, e.g., increasing 4 semitones on note C to get note E) and can still obtain high-quality singing voices.

\begin{table}[h] 
\begin{minipage}[t]{0.5\textwidth} 
\centering
\caption{CMOS for pitch and V/UV.}
\label{cmos3}
\begin{tabular}{lc}
\\ \toprule[1.5pt]
\textbf{System}      &   \textbf{CMOS}   \\ 
\midrule[1pt]
HiFiSinger  & 0  \\
\midrule
HiFiSinger without V/U input & -0.28 \\
HiFiSinger without F0 and V/U input & -0.34 \\
\bottomrule[1.5pt]
\end{tabular}
\end{minipage}
\begin{minipage}[t]{0.5\textwidth} 
\centering
\caption{CMOS under different window/hop sizes.}
\label{cmos4}
\begin{tabular}{lc}
\\ \toprule[1.5pt]
\textbf{System}                   & \textbf{CMOS}   \\ 
\midrule[1pt]
HiFiSinger (default, 20ms/5ms) & 0  \\
\midrule
HiFiSinger with 12ms/3ms & -0.36 \\
HiFiSinger with 50ms/12.5ms & -0.12 \\
\bottomrule[1.5pt]
\end{tabular}
\end{minipage}

\end{table}


\emph{Window/Hop Size} As analyzed in Section~\ref{sec_method_designs}, the window/hop size need be carefully chosen to consider the characteristics of singing voices as well as the trade-off of the model difficulty between acoustic model and vocoder. We study different window/hop size (we always set the ratio between window size and hop size to 4:1 following the common practice) in Table~\ref{cmos4}. It can be seen that larger or smaller window/hop size will cause quality drop, which demonstrates the effectiveness of our choice on window and hop size.

\begin{wraptable}{r}{0.50\textwidth}
\vspace{-0.6cm}
\centering
\small
\caption{CMOS under different receptive fields.}
\label{cmos5}
\begin{tabular}{lc}
\\ \toprule[1.5pt]
\textbf{System}                & CMOS   \\ 
\midrule[1pt]
HiFiSinger (default, 13 kernel size)                    & 0 \\
\midrule
HiFiSinger with 5 kernel size  & -0.39 \\
HiFiSinger with 9 kernel size  & -0.25 \\
\bottomrule[1.5pt]
\end{tabular}
\vspace{-0.1cm}
\end{wraptable}

\emph{Receptive Field} Unlike speaking voice, the duration in the music note and corresponding lyric may vary a lot, causing a large range of phoneme duration (usually longer and sometimes shorter), mainly on vowels. To better model the large range of duration, we use a larger kernel size in the vocoder to enlarge the receptive field to cover such long vowels. We conduct the CMOS evaluation on different sizes of convolution kernel. As shown in Table~\ref{cmos5}, a kernel size with larger receptive field can lead to improvement in audio quality.

\section{Conclusion}

In this paper, we have developed HiFiSinger, an SVS system to synthesize high-fidelity singing voice. To address the challenges caused by high sampling rate, we designed a SF-GAN on acoustic model to better model the wider frequency band, a ML-GAN on vocoder to better model longer waveform sequences, and introduced several systematic designs and findings that are important to improve singing modeling. Experiment results show that HFiSinger synthesizes singing voices with much higher quality than previous systems. For future work, we will continue to close the quality gap between the synthesized voices and recordings, and also apply our fidelity solution in HiFiSinger to text to speech synthesis.

\bibliography{iclr2021_conference}
\bibliographystyle{iclr2021_conference}


\end{document}